\documentclass[%
 aip,
rsi,%
 amsmath,amssymb,
 reprint,%
]{revtex4-1}

\usepackage{graphicx}
\usepackage[utf8]{inputenc}
\usepackage[T1]{fontenc}
\usepackage{dcolumn}
\usepackage{color}
\usepackage{bm}
\begin{document}

\preprint{AIP/123-QED}

\title[{\it Sahu et. al.}]{Progressive magnetic softening of ferromagnetic layers in multilayer ferromagnet-nonmagnet systems and the role of granularity}

\author{Siddharth~S.~Sahu}
 \affiliation{School of Physical Sciences, National Institute of Science Education and Research, HBNI, Jatni - 752050, India.}
 
\author{Vantari~Siva}
\affiliation{School of Physical Sciences, National Institute of Science Education and Research, HBNI, Jatni - 752050, India.}

\author{Paresh~C.~Pradhan}
\affiliation{Indus Synchrotrons Utilization Division, Raja Ramanna Centre for Advanced Technology, Indore 452013, India.}

\author{Maheswar~Nayak}
\affiliation{Indus Synchrotrons Utilization Division, Raja Ramanna Centre for Advanced Technology, Indore 452013, India.}
 
\author{Kartik~Senapati}
\email{kartik@niser.ac.in}
\affiliation{School of Physical Sciences, National Institute of Science Education and Research, HBNI, Jatni - 752050, India.}

\author{Pratap~K.~Sahoo}
\email{pratap.sahoo@niser.ac.in}
\affiliation{School of Physical Sciences, National Institute of Science Education and Research, HBNI, Jatni - 752050, India.}

\date{\today}

\begin{abstract}
We report a study of the structural and magnetic behavior of the topmost magnetic layer in a ferromagnet-nonmagnet (Co-Au) multilayer system. Glancing angle X-ray diffraction measurements performed on a series of multilayers showed a gradually increasing degree of amorphization of the topmost magnetic layer with increasing number of bilayers. Concurrently, the magnetic hardness and magneto-crystalline anisotropy of the top Co layer were found to decrease, as observed by magneto-optical Kerr effect measurements. This magnetic softening has been discussed in the light of Herzer’s random anisotropy model. Micromagnetic simulations of the multilayer system also corroborated these observations.
\end{abstract}

\maketitle

\section{Introduction}

Ferromagnet-nonmagnet multilayers have been extensively studied in the context of giant magnetoresistance (GMR) \cite{GMR}, data storage \cite{data-storage}, exchange interactions \cite{exchange-interaction} and several other applications\cite{thermoelectric,carlos2007,procspie}. Most of these effects are known to have a significant dependence on the interface morphology, among other parameters such as, thickness of individual components, and composition of the multilayers. Fundamentally, this interface dependence comes from variations in spin-dependent scattering of charge carriers at the interface. Therefore, issues like inter-diffusion and consequent compositional changes at the interface are some of the major bottlenecks in the field. This problem has been appreciated long back and several attempts have been made to optimize magnetoresistance in multilayers with immiscible components such as Co-Cu multilayers. GMR values of 5$\%$ have been measured at room temperature in granular Co-Cu films prepared by electro-deposition\cite{electroGMRCoCu}. People have also studied GMR in such multilayers as a function of thickness of Co layer \cite{CoGMR}, Cu layer \cite{CuGMR}, enhancement of GMR upon use of Ag additive on Cu layer\cite{AgAddGMR}, thermal effects in GMR \cite{GMTR} and applications in highly sensitive and flexible electronics\cite{flexible-electronics}.

As such, the magnetic properties of individual magnetic layers in multilayer films are dependent on magneto-crystalline anisotropy and exchange interactions, which dominate in the regime of higher and lower grain sizes, respectively. The crossover length for Co, which is one of the most widely used magnetic layer in multilayer structures, is $\sim$10 nm\cite{ferroexchange}. Typically, in thin films grown at ambient temperatures, grain sizes are known to depend on deposition rate and ambient pressure. However, the structural and morphological evolution of constituent magnetic layers, with increasing number of bilayers, have been largely ignored. Both theoretical\cite{theory1, theory2, theory3} and experimental\cite{expt0, expt1, expt2} studies on magnetic multilayers implicitly assume that the magnetic and structural properties are uniform across all the constituent layers in a multilayer system. In this report we have investigated this aspect of magnetic-nonmagnetic system in a multilayer of immiscible Co and Au, deposited using e-beam evaporation. Using glancing angle X-ray diffraction measurements we have followed the structural evolution of the top magnetic layer as a function of number of bilayers, which showed a gradually increasing degree of amorphization (reduction in grain size). Magneto-optical Kerr effect (MOKE)  measurements showed a concurrent magnetic softening of the top magnetic layer as the number of bilayers was increased. Micromagnetic simulations also supported our observation. 

\section{Experimental Details}
Si substrates were cleaned in acetone and then in isopropyl alcohol to remove surface impurities. 1, 2, 5 and 10 bilayers of Co and Au thin films were deposited on them using e-beam evaporation technique. A high vacuum chamber was used for this purpose with a base pressure below 1$\times 10^{-7}$ mbar. The thickness of each layer was $\sim$5 nm, as shown by quartz crystal monitor (QCM). The rates of deposition, 0.6 nm/min and 0.4 nm/min for Co and Au respectively, were maintained throughout the deposition for all the layers. X-ray reflectivity (XRR) measurements were performed on the bilayers to assess the roughness and to check the thickness of the individual layers. These parameters were extracted from the experimental data using Parrat's recursive algorithm in GenX package\cite{bjorck2007}. Evolving granularity of the multilayer films as a function of number of bilayers was studied using glancing angle X-ray diffraction method at an incident angle of 0.5 degrees to the surface. The magnetic behavior of the films were examined using Magneto-optical Kerr effect (MOKE) measurements at ambient temperatures. For all MOKE measurements, longitudinal mode was used with DC magnetic field aligned in the plane of the films.  

\section{Results} 
\begin{figure}[h]
 \centering
\includegraphics[scale=0.067]{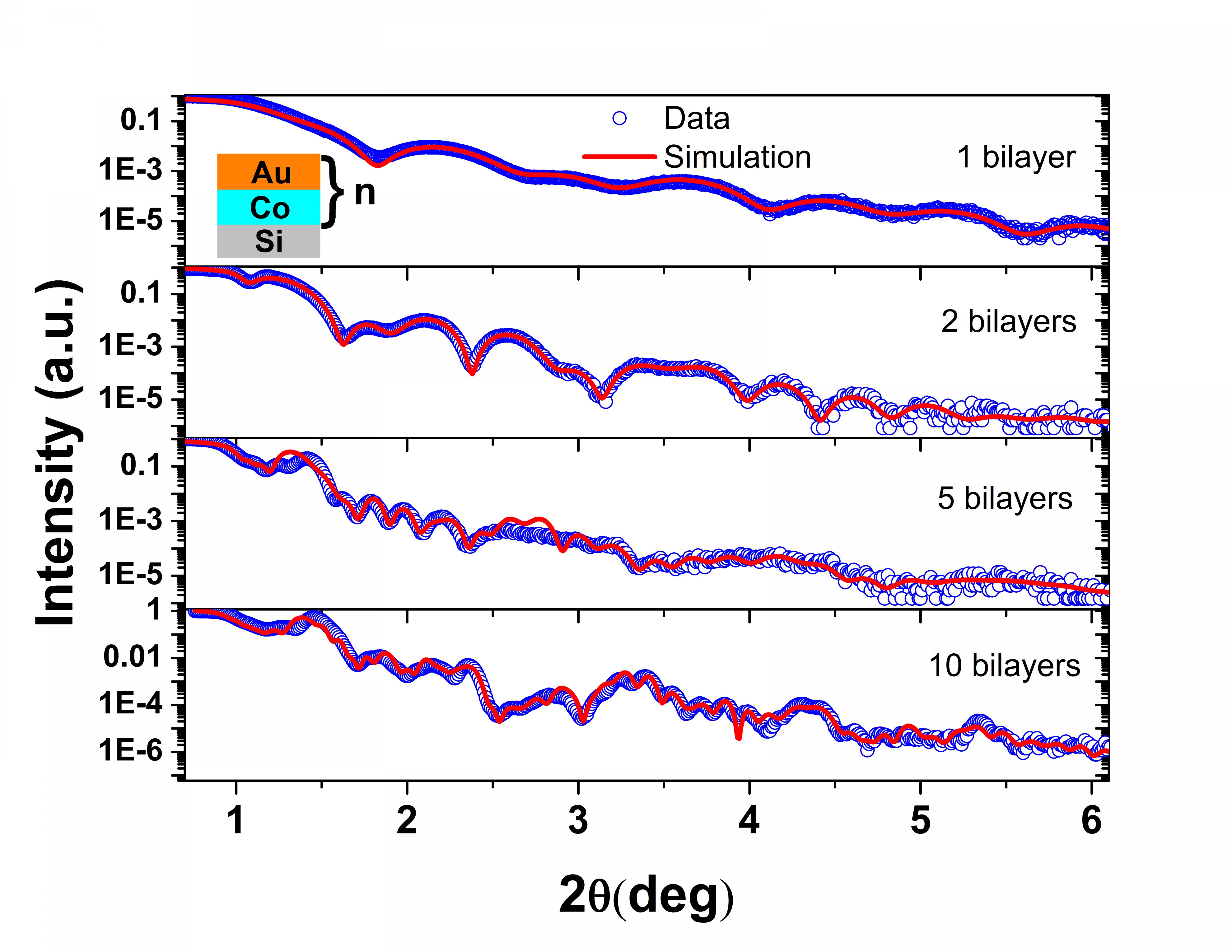}
\caption{X-ray reflectivity data of 1,2,5 and 10 bilayers of Co/Au along with the fitting curves. A schematic of the bilayers is shown in the inset.}
\label{XRR} 
\end{figure}

The XRR data along with the simulations are shown in Fig. \ref{XRR} and the average thickness and roughness of the layers in each sample and the errors in the parameters, obtained from fitting the data, are given in Table 1. The thickness values from the fitting are in good agreement with the values measured by QCM. The roughness of each layer is small and can be attributed to the immiscibility of Au and Co, which is due to positive heat of mixing ($\Delta H _{mix}=+11kJ/mol$)\cite{mixing}. In case of 5 and 10 bilayers the successive higher order Bragg peaks are not well defined. They are broad and show splitting because of presence of thickness error ($\sim12\%$) in the deposited layers. The top Au layer further serves to protect the underlying Co layers from oxidation.
 
\begin{table}[h]
\centering
\caption{Average thickness and roughness in nm obtained from XRR fitting. The subscript corresponds to the no. of bilayers.}
\label{table-thickness}
\begin{tabular}{ccc}
\hline\hline
Parameter       & Thickness & Roughness \\
\hline\hline
Au$_1$  & 6.42$\pm$0.15      & 0.39$\pm$0.09      \\
Co$_1$  & 4.79$\pm$0.15      & 0.60$\pm$0.09      \\
Au$_2$  & 6.23$\pm$0.15      & 0.62$\pm$0.09      \\
Co$_2$  & 4.07$\pm$0.15      & 0.69$\pm$0.09      \\
Au$_5$  & 5.06$\pm$0.15      & 0.74$\pm$0.09      \\
Co$_5$  & 4.75$\pm$0.15      & 0.81$\pm$0.09      \\
Au$_{10}$  & 4.46$\pm$0.15      & 0.69$\pm$0.09      \\
Co$_{10}$  & 4.34$\pm$0.15      & 0.72$\pm$0.09     \\
\hline
\end{tabular}
\end{table}

The immiscibility of Co and Au is also exhibited in the XRD data (in Fig. 2) which shows no alloy phases or compounds of Co and Au. The depth of penetration (d) of x-rays was calculated from the relation\cite{glxrd} $d=\sin \theta / \mu$, where the angle of incidence, $\theta=0.5^o$ and $\mu$ is the absorption coefficient of the material, which came out to be about 6$\pm$3 nm for Au and 25$\pm$5 nm for Co. Hence, the thicknesses of the Co and Au layers used are enough for the X-rays to probe the structure of the top bilayer, which is evidenced from the fact that we are able to obtain peaks in the XRD data for the case of 1 bilayer of Co and Au. One can then see that the peaks of metastable $\alpha$ Co (200) fcc and Co (200) fcc phases are lower for the case of 2 bilayers compared to the case of 1 bilayer, and vanish completely for the sample with 5 bilayers. We explain this on the basis of increase in degree of amorphisation (granularity) of the top Co layer as the number of bilayers increase. Furthermore, even though hcp is the most stable phase of Co, in the present study, only fcc phase is seen because of the low rate of evaporation of Co, as explained in our earlier work\cite{siva}. The peaks corresponding to Au are broad in all the cases signifying that all the layers of Au are amorphous. 

\begin{figure}[h]
\centering
\includegraphics[scale=0.32]{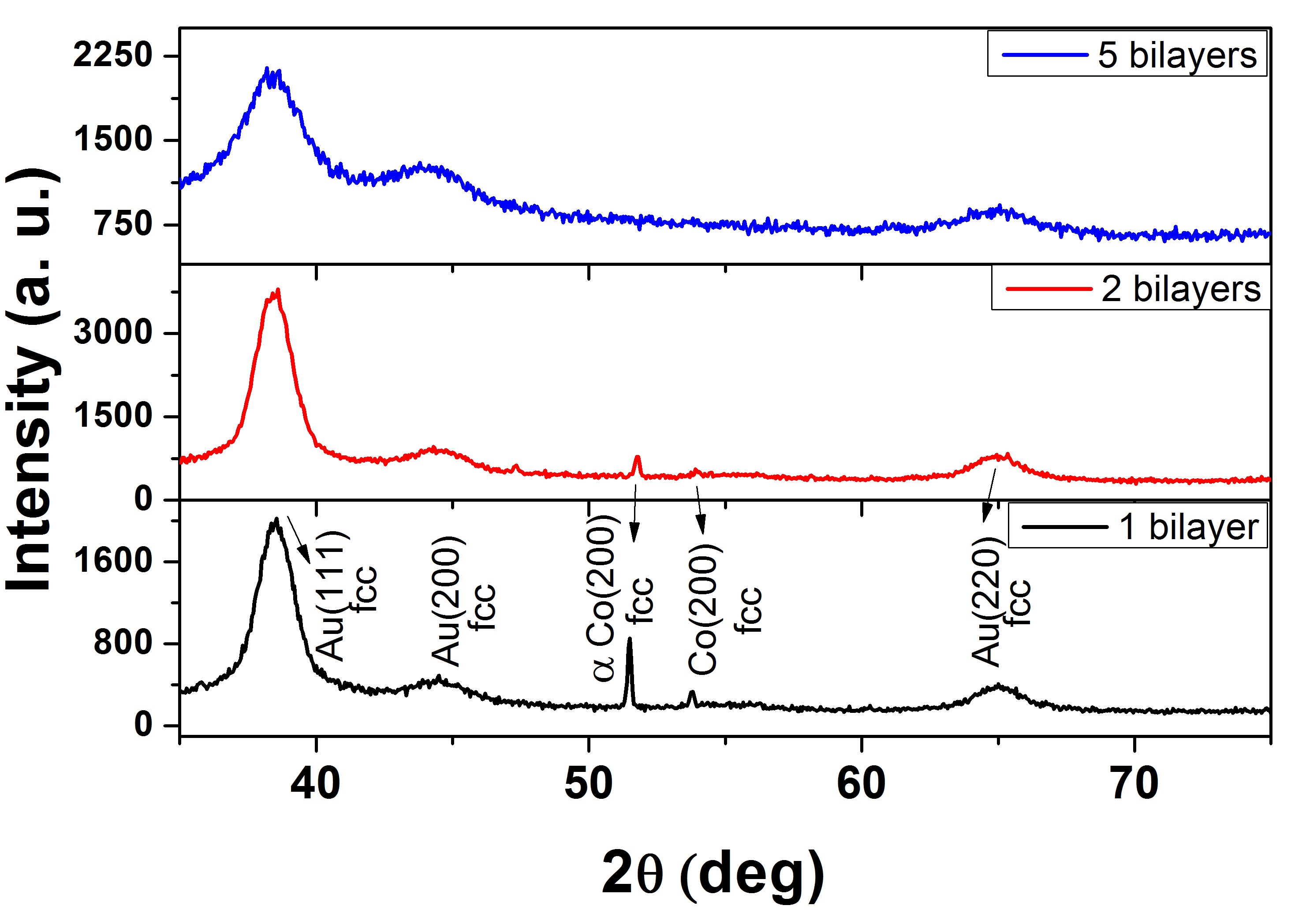}
\caption{Glancing angle XRD patterns of 1,2 and 5 bilayers of Co/Au showing increase in the amorphisation of the top Co layer as the number of bilayers is increased.}
\label{XRD} 
\end{figure}

The well-known Scherrer equation \cite{scherrer1918} was used to compute the average grain size, $D$ of Co from the FWHM of the XRD peaks:
\begin{equation}
D=\frac{K \lambda}{\beta \cos \theta}
\end{equation} 
 where $K$ is a shape dependent constant, $\lambda$ is wavelength of the X-ray source used, $\beta$ is FWHM of the peak in radians and $\theta$ corresponds to the angle of incidence. Using the values $K=0.9$ and $\lambda=1.5406$\AA $ $ the average grain sizes were calculated and are tabulated in Table 2. Since the Co peaks are nonexistent in the XRD data for the case of 5 bilayers, the grain sizes could not be calculated in that case. We must note that in case of X-rays incident at glancing angles, the application of Scherrer equation assumes that grains are like cylindrical plates \cite{cyl-plates} and the grain sizes thus calculated are permitted to be greater than the film thickness. In addition, we also emphasize the fact that due to the low glancing angles we are essentially probing the top Au and top Co layers only. 

\begin{figure}[h]
\centering
\includegraphics[scale=0.32]{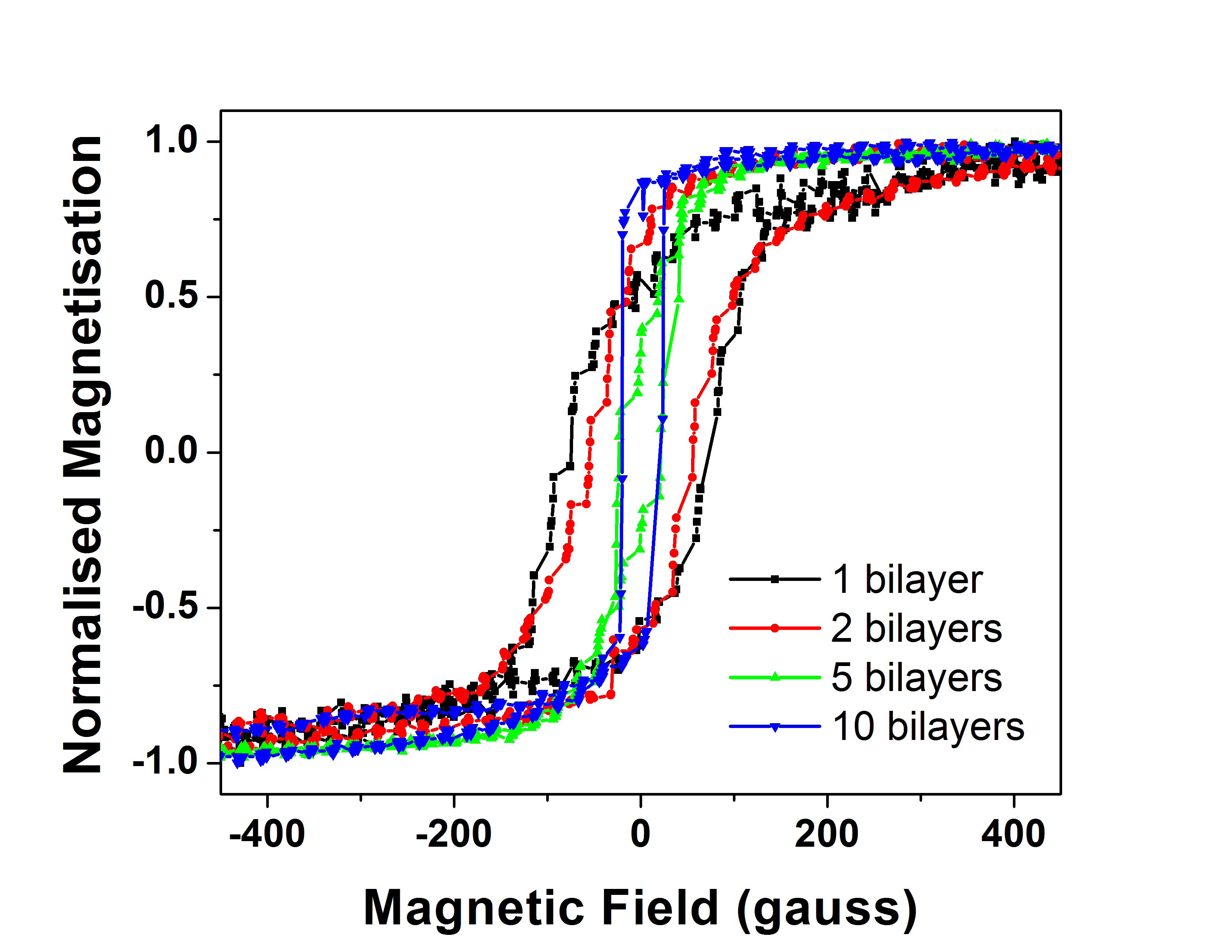}
\caption{Hysteresis loops obtained from MOKE measurements for 1,2,5 and 10 bilayers show that the coercivity decreases with the number of bilayers.}
\label{moke_1} 
\end{figure}

\begin{table}[h]
\centering
\caption{Grain sizes corresponding to the various peaks and number of bilayers (n) from the XRD data}
\label{grain_size}
\begin{tabular}{ccccc}
\hline\hline
Type            & n & $\beta$(deg) & $\theta$(deg) & $D$(nm) \\
\hline
$\alpha$ Co fcc & 1 & 0.140       & 25.744       & 62.729 \\
                & 2 & 0.183       & 25.883       & 48.273 \\
                \hline
Co fcc          & 1 & 0.133       & 26.889       & 67.021 \\
                & 2 & 0.193       & 26.971       & 46.074\\
\hline
\end{tabular}
\end{table}
 
We now discuss the correlation of magnetic property with the grain sizes. During the crystalline to amorphous transition of a material, the structural correlation length is reduced i.e. the grain size decreases. There exists a ferromagnetic exchange length, $L_{ex}$ below which the exchange energy starts to balance the  magneto-crystalline anisotropy energy of the grains, and is given by $L_{ex}=\sqrt{A/K_1}$, where $A$ is the exchange stiffness constant and $K_1$ is the anisotropy constant which is related to the crystal symmetry. When the average grain size D is of the order of the exchange length $L_{ex}$, it can then be used as a measure of the extent to which the exchange energy balances the magneto-crystalline anisotropy energy. As $D$ goes below $L_{ex}$, the directions of the magnetization and anisotropy direction of the individual grains start to become increasingly uncorrelated. While the magnetization directions of the grains are progressively aligned parallel to the field, the anisotropy direction can remain randomly oriented, since the torque on them due to the field is reduced significantly by exchange interaction with other grains. As a result, the effective anisotropy, averaged over several grains, is considerably reduced in magnitude. Using statistics and scaling arguments the random anisotropy model\cite{herzer1989,herzer1990} derives a relationship between the effective magneto-crystalline anisotropy constant $\langle K_1\rangle$ and the average grain size $D$ . The model assumes that an assembly of exchange coupled grains, each of size $D$ are embedded in an ideally soft ferromagnetic matrix, occupying a volume fraction $v_r$ and having their magneto-crystalline anisotropies oriented randomly. $L_{ex}$ determines the ferromagnetic correlation volume $V=L_{ex}^3$ which contains $N$ number of grains over
 
 \begin{figure}[hbtp!]
\centering
\includegraphics[scale=0.36]{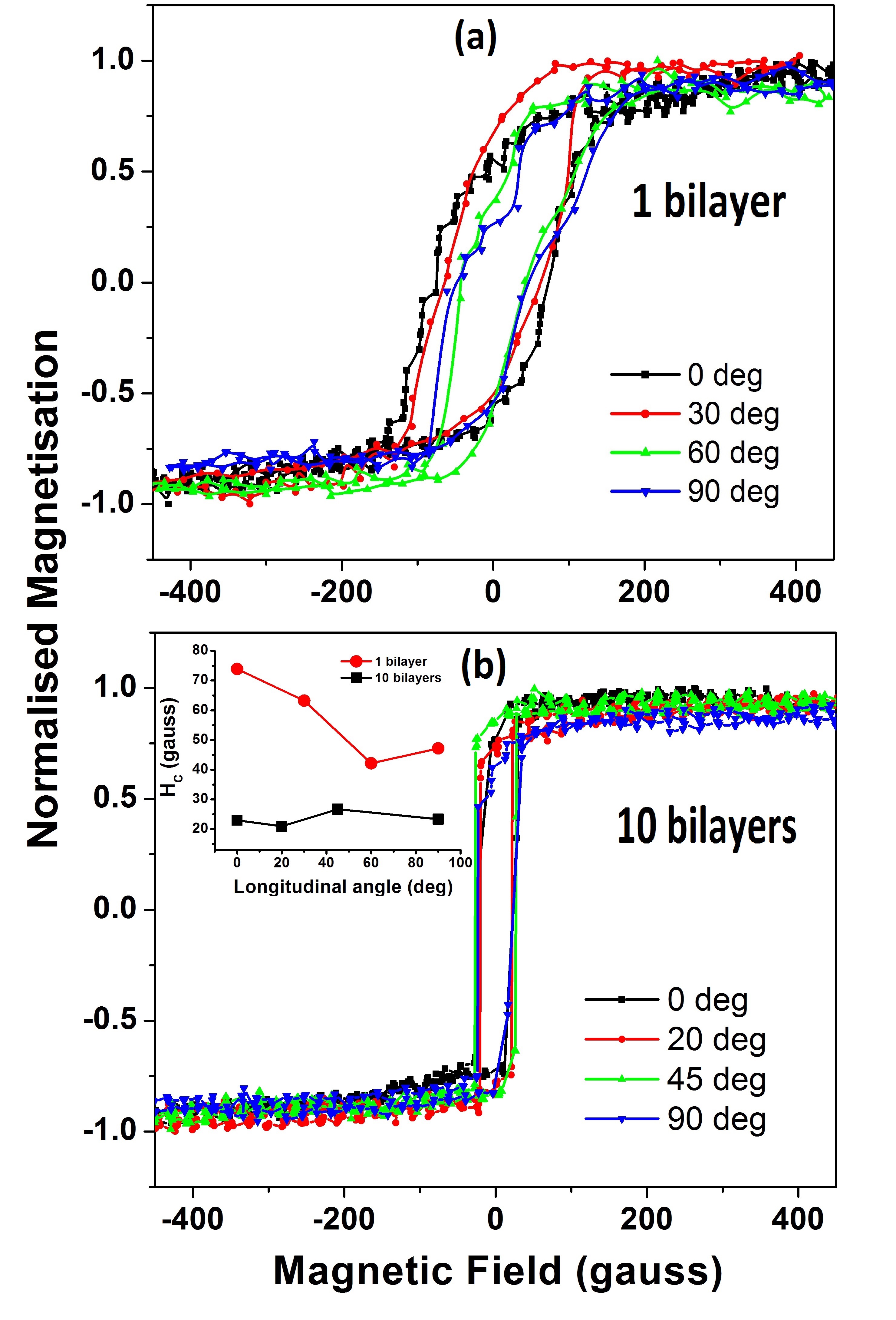}
\caption{MOKE data of (a) 1 bilayer and (b) 10 bilayers at various angles of the sample with the longitudinal axis. Inset shows plot the coercivity H$_C$ with the longitudinal angle for 1 and 10 bilayers.}
\label{moke_2} 
\end{figure}
 \begin{figure}[hbtp!]
\centering
\includegraphics[scale=0.34]{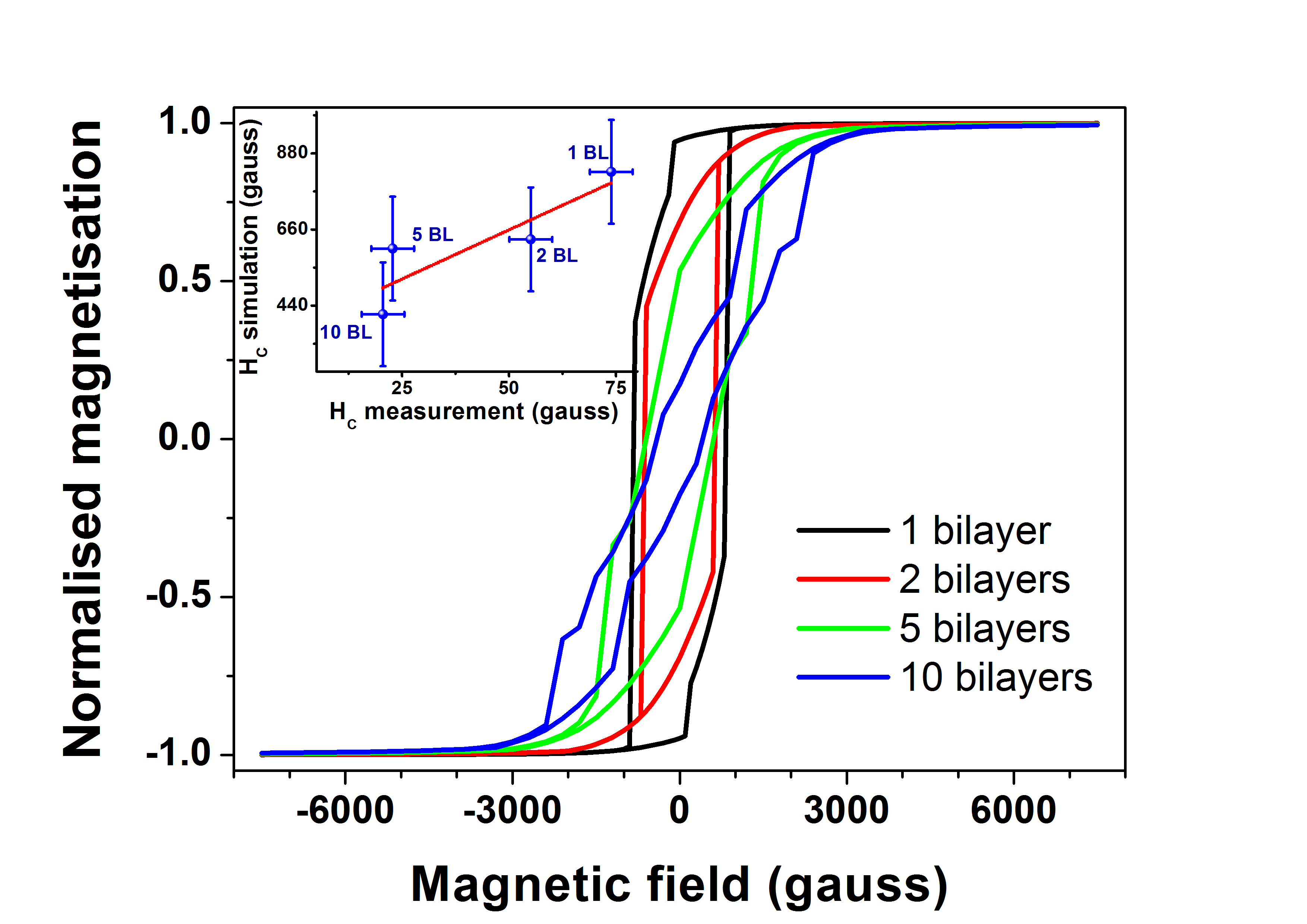}
\caption{Simulated hysteresis curves of the 1,2,5 and 10 bilayers through micromagnetic simulations using OOMMF. Inset shows plot of simulated H$_{C}$ versus experimental H$_{C}$ for different bilayers (BL).}
\label{oommf_1} 
\end{figure}
 \begin{figure*}[hbtp!]
\centering
\includegraphics[scale=0.41]{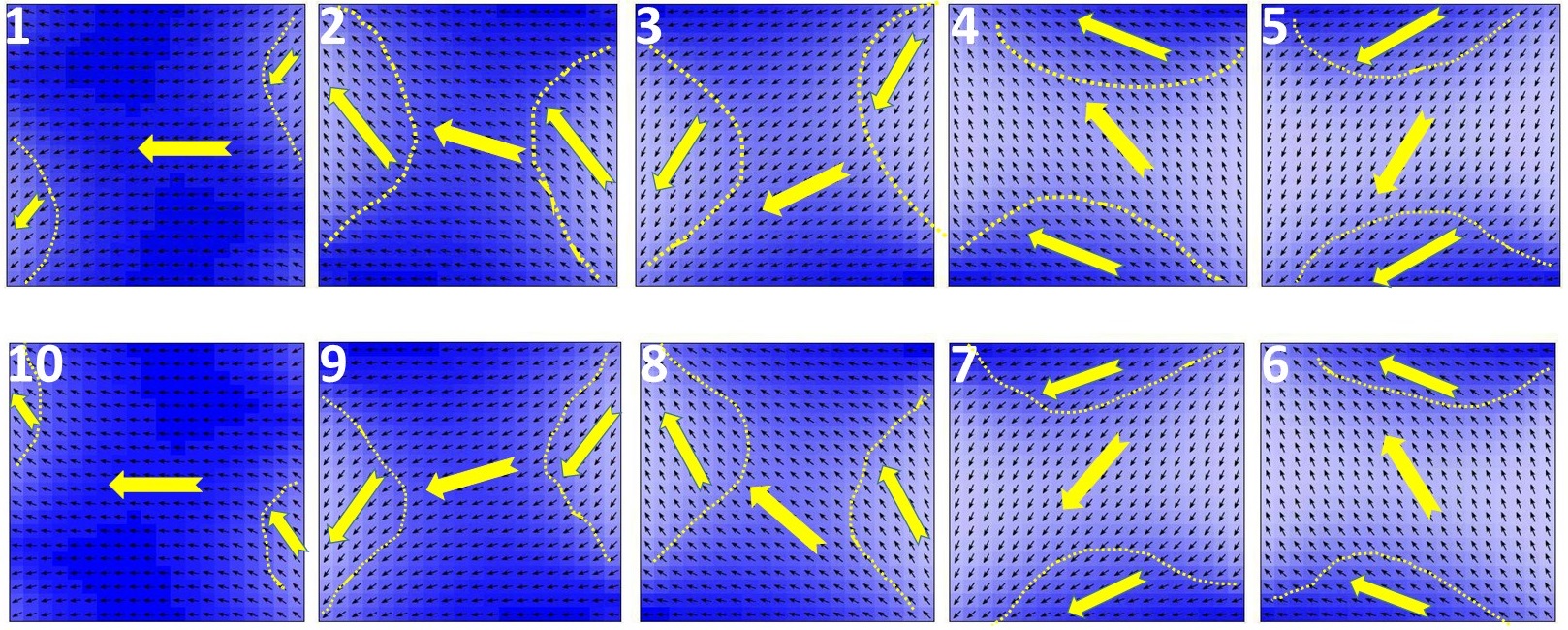}
\caption{ Spin orientations of all the Co layers of the 10 bilayers sample near the saturation field through OOMMF simulations. Color scheme for the spins orientation: blue - along the field, red - opposite to the field and white - intermediate directions. The yellow arrows indicate the general orientation of the spins in a region.}
\label{oommf_2} 
\end{figure*}
\noindent which we take the average to arrive at the effective anisotropy constant $\langle K_1\rangle$. We note that $N=v_r(L_{ex}/D)^3$ and that statistical fluctuations for some finite $N$ will lead to the existence of a direction with minimum energy because of which the effective anisotropy energy density can then be determined from the mean of the fluctuation amplitude as:
 \begin{equation}
 \label{K}
 \langle K \rangle = \frac {v_r K_1}{\sqrt{N}} = v_r K_1 \left( \frac{D}{L_{ex}}\right)^{3/2}
 \end{equation}

Scaling $L_{ex}$ by replacing $K_1$ by $\langle K \rangle$, we get:
 \begin{equation}
 \label{Lex}
 L_{eq}=\sqrt{A/\langle K \rangle}
 \end{equation} 
Combining equations $\ref{K}$ and $\ref{Lex}$, we finally arrive at the expression relating $\langle K \rangle$ and $D$
 \begin{equation}
 \langle K \rangle = v_r^2 K_1^4 D^6 /A^3
 \end{equation}
 In the absence of any other forms of anisotropy, the coercivity $H_c$ which is proportional to $\langle K \rangle$ is hence shown to vary with the 6$^{th}$ power of the grain size, $D$ when it is below the exchange length. We showed earlier using the Scherrer equation from the XRD data that the grain size showed a decreasing trend and for higher number of bilayers, we expect it to be close to the ferromagnetic exchange length\cite{ferroexchange} of Co $\sim$ 10nm (but this cannot be confirmed from the present data due to absence of XRD peaks) and hence its magneto-crystalline anisotropy constant reduces significantly with increase in the number of bilayers.

 In order to investigate the relation between crystal structure and magnetic properties, we have performed MOKE measurements on the series of samples with varying number of bilayers. In Fig. \ref{moke_1}, we plot the normalized magnetization of the multilayers as a function of magnetic field. A clear decrease in the coercive field was observed with increasing number of bilayers. We emphasize here that for the MOKE measurements, a 633 nm (red) laser was used, which has a characteristic penetration depth of $\sim$10 nm\cite{depthAu}. Therefore, the observed magnetic softening of the Kerr loops in Fig. \ref{moke_1} corresponds to the top Co layer only. This is consistent with the discussion in the previous section. It is known that coupling between electron spin and crystal field leads to a directional asymmetry (anisotropy) in magnetic properties. 

As smaller grains tend to be influenced more by exchange interaction, anisotropy directions remain randomly oriented even under applied field. This effect was verified by angle dependent MOKE measurements, where direction of the applied magnetic field was changed in the plane of the multilayers. Panels (a) and (b) in Fig. \ref{moke_2} show the angular variation of magnetization loops for 1 bilayer and 10 bilayers, respectively. The coercive fields extracted from these measurements are plotted in the inset of panel (b) in Fig. \ref{moke_2}. We notice that there is hardly any angular dependence in the case of 10 bilayers sample whereas 1 bilayer sample shows a clear anisotropy. This observation adds further support to the relevance of the random anisotropy model in our case. 

To further understand the magnetization dynamics, micromagnetic simulations were carried out using the OOMMF package\cite{oommf}. The thickness of each layer was 5nm, cell size was 5 nm$\times$5 nm$\times$1 nm and the simulation was carried out in a 100 nm$\times$100 nm square geometry. The magnetic parameters\cite{oommf_parameters} used for the simulations were saturation magnetization M$_s$ = 1400 kA/m, exchange stiffness A = $1.12\times 10^{-11}$ J/m and anisotropy constant K$_1$ = $2.15\times 10^{6}$ J/m. As justified earlier, the anisotropy constant was progressively reduced for increasing number of bilayers in the simulation. Fig. \ref{oommf_1} shows the simulated hysteresis curves for the 1, 2, 5, and 10 bilayer samples. The coercive fields extracted from the simulations are plotted with the experimentally observed coercive fields, in the inset of Fig. \ref{oommf_1}. They appear to follow a linear relation within errors. The saturation fields observed in Fig \ref{oommf_1}, however, show an opposite trend as a function of the number of bilayers, when compared to the experimental data shown in Fig \ref{oommf_1}. In order to understand this apparent discrepancy, in Fig \ref{oommf_2} we have shown the simulated magnetic configuration of all magnetic layers in a 10 bilayer system at 2700 gauss, which is close its saturation field, 3300 gauss. Clearly, at this field, the top and bottom magnetic layers have achieved almost complete saturation while the intermediate layers are far from saturation. Since the hysteresis loops obtained in the simulations are averaged over all the layers, the apparent saturation field is higher. On the other hand, since the MOKE measurements are actually probing only the top magnetic layer, the saturation field is much less. 

\section{Conclusions}
In summary, we have investigated the structural evolution and the consequent changes in magnetization dynamics of magnetic layers in ferromagnet-nonmagnet multilayer systems. Using glancing angle X-ray diffraction we were able to follow the structural changes of the top magnetic layers in multilayer systems. Low penetration depth of the light source in magneto-optical measurements provided the required resolution for following the magnetization evolution of the top magnetic layer only. Contrary to the general belief, a clear gradual change was observed in both structural and magnetic properties of the successive magnetic layers. An increasing degree of amorpization (grain size reduction) of the top layer with increasing number of bilayers resulted in a significant reduction in magnetic hardness of the top layer. The magnetic softening was discussed in the light of random anisotropy model. This observation was also found consistent with micromagnetic simulations.  

\section{Acknowledgements}
We acknowledge the funding from National Institute of Science Education and Research (NISER), Department of Atomic Energy (DAE), India. We would also like to thank S. K. Rai, S. Prusty and  S. S. Mishra for XRR, MOKE measurements and OOMMF simulations respectively.

\end{document}